\documentclass[namedreferences]{solarphysics}
\usepackage[hyperref,optionalrh,solaromanenum]{spr-sola-addons} 
\usepackage{graphicx}                    
\usepackage{color}                       
\usepackage{breakurl}                         

\begin{document}

\begin{article}

\begin{opening}

\title{Ion Charge States in a Time-Dependent Wave-Turbulence-Driven Model of the Solar Wind}

%
\author[addressref={1},corref,email={lionel@predsci.com}]{\inits{R.}
\fnm{Roberto}~\lnm{Lionello}\orcid{0000-0001-9231-045X}}
\author[addressref={1}]{\inits{C.}\fnm{Cooper}~\lnm{Downs}}
\author[addressref={1}]{\inits{J.A.}\fnm{Jon A.}~\lnm{Linker}}
\author[addressref={1}]{\inits{Z.}\fnm{Zoran}~\lnm{Miki\'c}}
\author[addressref={2}]{\inits{J.}\fnm{John}~\lnm{Raymond}}
\author[addressref={2}]{\inits{C.}\fnm{Chengcai}~\lnm{Shen}}
\author[addressref={3}]{\inits{M.}\fnm{Marco}~\lnm{Velli}}

%

\address[id={1}]{Predictive Science Inc., 9990 Mesa Rim Rd.\, Suite 170, San
Diego, CA 92121, USA}
\address[id={2}]{Harvard-Smithsonian Center for Astrophysics, 60 Garden St., Cambridge, MA 02138, USA}
\address[id={3}]{Earth, Planetary, and Space Sciences, UCLA, 595 Charles Young Dr. E., Box 951567 Los Angeles, CA 90095, USA}

\begin{abstract}
Ion fractional charge states, measured  \textit{in situ} in the heliosphere,    
depend on the properties of the plasma in the inner corona.
As the ions travel outward in the solar wind and the electron density drops, 
the charge states remain essentially  unaltered or ``frozen in".
Thus they can provide
a powerful constraint on  heating  
models
 of the corona and acceleration of the solar wind. We have implemented
non-equilibrium ionization calculations into a 1D 
wave-turbulence-driven (WTD) hydrodynamic solar wind model
 and
compared modeled charge states with the Ulysses 1994-5 \textit{in situ}
 measurements.
We have found that modeled charge state ratios of
 $C^{6+}/C^{5+}$ and  $O^{7+}/O^{6+}$,
 among others, were too low compared with Ulysses measurements. 
However, a heuristic reduction of the plasma flow speed 
has been able to bring the modeled results in line with observations,
 though other ideas have been proposed to address this discrepancy.
We discuss implications of our results and the 
prospect of including ion charge
state calculations into our 3D MHD model of the inner heliosphere.
\end{abstract}

%

\end{opening}

%
\section{Introduction}

Fractional charge states of ions in the solar corona are determined by
the local properties of the plasma. However, 
the rapidly decreasing electron density
of the plasma released into the solar wind
prevents further ionization and recombination beyond a few solar radii 
\citep[][and references therein]{10.1029/1999RG000063,2002SSRv..101..229C}.
Thus measurements
of charge states in the heliosphere such as those performed by Ulysses/SWICS \citep{2002GeoRL..29.1352Z} give us information and constraints on the
properties of the corona from which they originate, 
the higher-ionization states
being associated with hotter regions.
While ionization equilibrium is a valid assumption in many
cases and especially in the lower corona, it does not apply when 
the dynamic time scales of the plasma
are shorter than those of  ionization and recombination.
In those instances   charge states must be calculated with
  a time-dependent scheme, which is then relaxed to a steady state 
for the steady solar wind solutions computed here
\citep{2015A&C....12....1S}.
Although the evolution of the charge state distribution in the solar wind 
has been studied with sophisticated multi-fluid models
\citep{1986SoPh..103..347B,1998ApJ...498..448E,1999JGR...10417005K,2003ApJ...582..467C,2011ApJ...732..119B}, connecting  \textit{in situ} measurements with coronal spectroscopic data
still remains problematic \citep{2014ApJ...790..111L}. 
This very difficulty makes the reproduction of charge-states in MHD
 computational models of the solar corona a robust    constraint 
for the validation of the models themselves.
{
\citet{2015ApJ...806...55O}  pioneered in this  effort by using
an external code to evaluate 
charge states in  the solar wind calculation obtained 
with the Alfv\'en Wave Solar Model (AWSoM), and comparing them with   \textit{in situ}
measurements from Ulysses.
}
Recently we incorporated a Wave-Turbulence-Driven (WTD) 
formulation for coronal heating and solar wind acceleration by 
Alfv\'enic turbulence into a 3D MHD model of the global solar 
corona \citep{2018NatAs.tmp..120M}. In this effort, we constrained the 
model by forward modeling EUV, X-Ray, and white-light coronal emission 
and comparing directly to observations.
Although it is our long-term goal
to use the calculation of fractional charge states to 
further constrain our 3D model,
 it is expedient to start this process with our
1D solar wind model, since it contains analogous
heating and acceleration schemes 
\citep{2014ApJ...784..120L,2014ApJ...796..111L}.
With this aim in mind,
we have added the fractional charge states module of 
\citealp{2015A&C....12....1S}  to our 1D model. Then
  we have compared 
 the calculated 
 $C^{6+}/C^{5+}$ and  $O^{7+}/O^{6+}$ ratios and the average iron charge 
state, $<Q>Fe$
with those measured by Ulysses during 1994-5, when the spacecraft
spanned a large latitudinal interval.
Since our preliminary
 results  could not match the \textit{in situ} data, we have developed and
validated a  heuristic method to correct 
 the 1D model and improve the comparison with satellite measurements.
This modification
 can also be implemented in 3D calculations. This paper is organized as follows: in Sec.~\ref{s:model} we describe the 1D model;
the first results, the modifications to the model, and the
corrected results are in 
 Sec.~\ref{s:results}; we draw our conclusions in Sec.~\ref{s:conclusion}. 
\section{Model Description}\label{s:model} 
We use the
1D hydrodynamic (HD) model of the solar wind of 
\citealp{2014ApJ...784..120L,2014ApJ...796..111L}, which
is based on our  WTD 
formulation. This formulation uses the propagation, reflection, 
and non-linear dissipation of Alfv\'enic turbulence to heat and 
accelerate the solar wind \citep{2010ApJ...708L.116V}. 
This  model solves  the following set of time-dependent, 1D HD equations:
\begin{eqnarray}
\frac{\partial \rho}{\partial t}
  &=&  
-  \frac{1}{A}\frac{\partial}{\partial s}\left ( A U \rho \right ),  \label{eq-rho}
  \\
\rho  \frac{\partial U}{\partial t} &=& 
- \rho  U\frac{\partial U}{\partial s} - \frac{\partial }{\partial s}(p+p_w) + g_s \rho
+ \mathsf{R}_s  
 + \frac{1}{s^2}\frac{\partial}{\partial s} 
\left (s^2 \nu  \rho \frac{\partial U}{\partial s} \right ),  \label{eq-mom}
 \\
\frac{\partial T}{\partial t}
 &=& -U\frac{\partial T}{\partial s} \nonumber \\
 &-& (\gamma -1)\left [ T {\frac{1}{A}\frac{\partial}{\partial s}A} U 
 -\frac{m_p}{2 k \rho}
   \left( {\frac{1}{A}\frac{\partial}{\partial s}A} q 
    -  n_en_p{Q(T)}+{H}\right) \right],
\label{eq-T} 
\\
\frac{\partial z_\pm}{\partial t}
&=&
-[U\pm V_a] \frac{\partial z_\pm}{\partial s} -\frac{1}{2}[U\mp V_a]\left ( \frac{\partial \log V_a}{\partial s} 
\frac{\partial \log A}{\partial s} \right )z_\pm 
\nonumber \\
&+& \frac{1}{2}[U\mp V_a]\frac{\partial \log V_a}{\partial s} 
z_\mp - \frac{|z_\mp| z_\pm}{2\lambda_\odot \sqrt{A/A_\odot}
}, \label{eq-zpm}
\\
H&=&  \rho  \frac{|z_-| z_+^2+ |z_+| z_-^2}{ 4 \lambda_\odot \sqrt{A/A_\odot}} ,
 \\
p&=& 2 n k T, \\
p_w&=& \frac{1}{2}\rho \frac{ (z_- -z_+)^2}{8}, \\
\mathsf{R}_s&=& \rho z_+ z_- \frac{\partial \log A}{\partial s},
\end{eqnarray}
where
$s \geq  R_\odot$ is  the distance along a  magnetic field line; $p$, $T$,  $U$, and $\rho$, are the plasma pressure, temperature, velocity,
and density. The number density, $n$, is assumed to be equal for protons ($n_p$)
 and electrons ($n_e$).
$k$ is Boltzmann constant.
$g_s=g_0 R_\odot^2 \mathbf{\hat{b}\cdot \hat{r}}/r^2$ is 
the gravitational acceleration parallel to the magnetic field line
($\mathbf{\hat{b}}$).  The  kinematic viscosity is  $\nu$.   $A(s)=1/B(s)$
is the area factor along the field line and  the inverse of the magnetic field magnitude $B(s)$.
The field aligned component of the 
vector divergence of the MHD Reynolds stress, 
$\mathbf{R}=(\delta \mathbf{b}  \delta \mathbf{b}/ 4 \pi - \rho \delta
\mathbf{u}  \delta \mathbf{u} )$, is $R_\mathsf{s}$. $\delta\mathbf{u}$ and  $ \delta\mathbf{b}$ are
 respectively the fluctuations of the 
velocity $\mathbf{u}=U(s) \mathbf{\hat{b}} + \delta \mathbf{u}$ and of the
magnetic field, $\mathbf{B}=B(s) \mathbf{\hat{b}} +\delta  \mathbf{b}$, 
with  $\mathbf{\hat{b}} \cdot \delta \mathbf{b} = 0= \mathbf{\hat{b}} \cdot \delta \mathbf{u}$. 
$ p_w = {\delta \mathbf{b}^2}/{8 \pi}  $ is the
wave pressure.  In  
Eq.~(\ref{eq-T}), the polytropic index is
 $\gamma=5/3$.  The radiation loss function $Q(T)$ is as in \citet{1986ApJ...308..975A}. 
For the heat flux $q$, according to the radial distance, 
either a
collisional (Spitzer's law) or collisionless form \citep{1978RvGSP..16..689H} is employed.
At a distance of $10R_\odot$ from the Sun, a smooth transition between the two forms  occurs 
\citep{1999PhPl....6.2217M}.
In Eq.~(\ref{eq-zpm}),
the Elsasser variables
$\mathbf{z}_\pm=\delta \mathbf{u}
\mp \delta \mathbf{b}/\sqrt{4 \pi \rho} $  \citep{2001ApJ...548..482D} are advanced. 
$\mathbf{z}_+$  represents an outward propagating perturbation along a radially outward 
 magnetic field line, while $\mathbf{z}_-$ is directed inwardly.
 The 
actual direction of $\mathbf{z}_\pm$ is assumed to be unimportant, provided that  it is in
the plane perpendicular to $\mathbf{\hat{b}}$ and that only low-frequency
perturbations are relevant for the heating and acceleration of the plasma. 
Hence, we treat $z_\pm$ as scalars. The Alfv\'en speed along the field line is
$V_a(s)=B/\sqrt{4 \pi \rho}$.  With 
$R_1^\pm $ and $R_2^\pm$ respectively, we indicate  the WKB and reflection terms, 
which are related to the large scale
gradients. $\lambda_\odot$ is the turbulence correlation scale at the solar surface.
Thus the heating function $H$ \citep{1938RSPSA.164..192D,2004GeoRL..3112803M},
$p_w$ and $\mathrm{R}_s$ 
\citep{2011ApJ...727...84U,2012ApJ...754...40U}
can all be expressed in terms of   $z_\pm$.
{We are allowed to specify temperature and  density at the lower  boundary because the solar wind is subsonic there. However, the velocity must
be  determined by solving 
the 1D gas characteristic equations.  }
Since the upper  boundary is placed
beyond all critical points,  the characteristic equations are used for all variables.
The amplitude of
the outward-propagating (from the Sun) wave is imposed in the $z_\pm$ equations.

\citealp{2014ApJ...784..120L} used the model to
 explore 
the parameter space of $\lambda_\odot$ and  $z_+^\odot$  ($z_+$ at the solar surface)  in a radial field line to  determine
the plasma speed, density, and temperature at 1 A.U. 
\citealp{2014ApJ...796..111L} calculated instead
 solar wind solutions at different latitudes along 
open flux tubes of the magnetic  field described in \citealp{1998A&A...337..940B}.
In the present work, in parallel with the HD equations, we use $U$, $T$, and $n_e$ to 
evolve the fractional
charge states of minor ions according to the model of  \citet{2015A&C....12....1S}:
\begin{equation}
\frac{\partial {}_{Z}F^i}{\partial t} + U \frac{\partial}{\partial s} {}_{Z}F^i 
=
n_e\left [ {}_{Z}C^{i-1} {}_{Z}F^{i-1} - \left ( {}_{Z}C^i +{}_{Z}R^{i-1} \right
 ) {}_{Z}F^i + {}_{Z}R^{i}  {}_{Z}F^{i+1} \right ]. \label{e:cs}
\end{equation}
For an element with atomic number $Z$, ${}_{Z}F^i(s)$ indicates the
fraction of ion ${i+}$ ($i=0,Z$) in respect of the total at a grid point:
\begin{equation}
\sum_{i=0}^{Z} {}_{Z}F^i = 1.
\end{equation}
For each element, the ion fractions are coupled through the ionization,
 ${}_{Z}C^{i}$(T), and recombination, ${}_{Z}R^{i}$(T), rate coefficients
derived
from the CHIANTI (version~7.1) atomic database
\citep{1997A&AS..125..149D,2013ApJ...763...86L}. Although in principle the values of the ion fractions 
could be used to determine the radiation law function $Q(T)$ in Eq.~(\ref{eq-T}), they provide
  no feedback effects in this investigation.  As initial condition, we prescribe at each point 
the equilibrium values of each ${}_{Z}F^i(s)$, which is obtained from  the 
module  of \citet{2015A&C....12....1S}.
{ As boundary condition at $s=R_\odot$  we keep
 the initial, equilibrium  ${}_{Z}F^i(R_\odot)$. At the outer boundary 
$s=215 R_\odot$, since the charge states are frozen-in,  we
set the values  to be the same as those at the grid 
point immediately preceding, ${}_{Z}F^i(215 R_\odot)={}_{Z}F^i(215 R_\odot -\Delta r)$. 
}
\section{Results}\label{s:results} 
We calculate the fractional charge states  in a parameter space study of the  fast solar wind 
and  for  the magnetic
field configuration of \citet{1998A&A...337..940B}. Since in either case the computed
ion fractions do not match  \textit{in situ} measurements, we devise a correction for the ion outflow speed. Then we show the 
calculated charge states with the corrected flow.
\subsection{Charge-States in a  Parameter Study of the Fast Solar Wind} \label{ss:fastwind}
 \begin{figure} 
 \centerline{\includegraphics[width=\textwidth,clip]{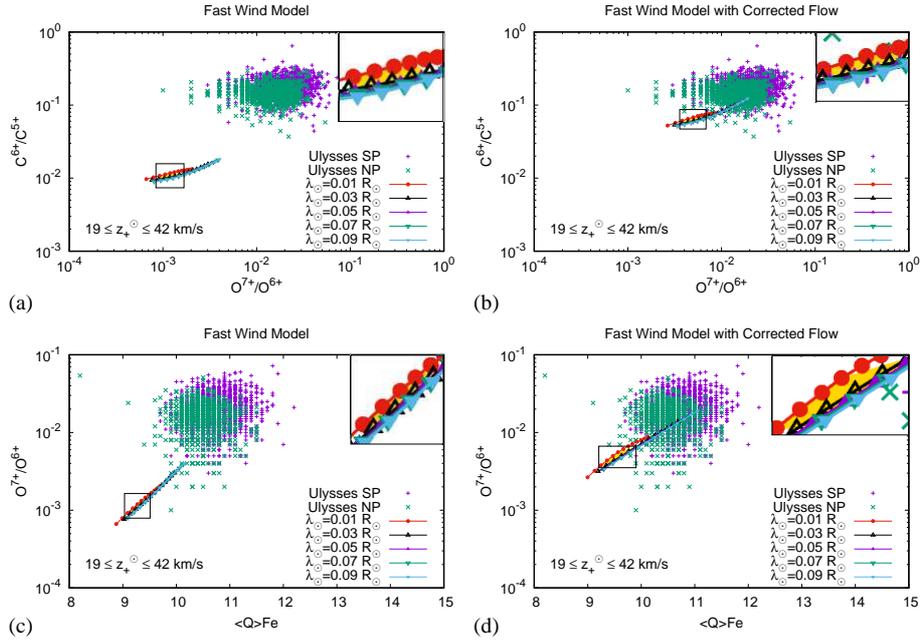}}
 \caption{The ion charge states at 1 A.U.\ in the parameter study of the solar wind in WTD model of 
\citet{2014ApJ...784..120L} compared with the measurements of Ulysses at latitudes $|\phi| \geq  70^\circ $
during 1994-5. Values from simulations with the same $\lambda_\odot$  are grouped along curves. Along each, curve
a symbol indicates the calculated result. $z_+^\odot$ increases from bottom left to top right {at 13 values equally spaced
between $19~\mathrm{km/s} \lesssim z_+^\odot \lesssim 42~\mathrm{km/s}$, 
the interval between each value being 
$\Delta z_+^\odot \simeq 1.9~\mathrm{km/s}$.} 
The thin area in
gold corresponds to solutions with  $630 \lesssim U \lesssim 820 \ \mathrm{km/s}$ and 
$1.5 \lesssim n_e \lesssim 3 \ \mathrm{cm^{-3}}$ for the plasma at 1 A.U.
{Enlargements around the areas are provided in the upper right corners
of each panel}.
(a)  $O^{7+}/O^{6+}$ vs.\ $C^{6+}/C^{5+}$ in the original model.  (b) The same as (a) with corrected flow in the evolution
of the ions.
(c) $<Q>Fe$  vs.\ $O^{7+}/O^{6+}$  in the original model. (d) The same as (c)  with corrected flow in the evolution
of the ions.
}\label{fig:fastwind}
 \end{figure}
Using the WTD model described in Sec.~\ref{s:model},
\citet{2014ApJ...784..120L} performed a parameter study of  the fast solar wind along a radial magnetic
 field line. They varied $\lambda_\odot$ at 5  values  within
 $0.01~R_\odot\leq \lambda_\odot \leq 0.09~R_\odot$,
 with an interval $\Delta \lambda_\odot=0.02~R_\odot$, and 
$z_+^\odot$  at 13 values equally spaced
between $19~\mathrm{km/s} \lesssim z_+^\odot \lesssim 42~\mathrm{km/s}$, 
the interval between each value being 
$\Delta z_+^\odot \simeq 1.9~\mathrm{km/s}$. 
Not all values yielded steady-state solutions: 
when $\lambda_\odot=0.01~R_\odot$, acceptable  
solutions were found only for $19~\mathrm{km/s} \lesssim z_+^\odot \lesssim
31~\mathrm{km/s}$; when $\lambda_\odot=0.03~R_\odot$, a
steady-state
solution was not found for $z_+^\odot \simeq 42~\mathrm{km/s}$.

 We have repeated
the same simulations, having activated  the ion charge states  evolution module for carbon, oxygen,  and iron.
In Fig.~\ref{fig:fastwind} we show comparisons between results of the computation and
 the  measurements of Ulysses/SWICS
\citep{2002GeoRL..29.1352Z}
during the years 1994 and 1995, when the spacecraft performed the rapid latitude scans.  Since the parameter
 study concerns
the fast solar wind, we  show measurements only for latitudes larger than
 $70^\circ $ north or south. Each symbol along the curves represents the results of solutions with the same  $\lambda_\odot$ 
but increasing
 $z_+^\odot$  from bottom-left to top-right.  Panel (a) has the ratio of $O^{7+}/O^{6+}$ on 
the $x$-axis and  $C^{6+}/C^{5+}$ on the the $y$-axis and Panel (c) has the average charge state of iron, $<Q>Fe$, versus
the  $O^{7+}/O^{6+}$ ratio [Panels (b) and (d) will be described later]. From Panel (a) it is evident
that, although in some instances values of $O^{7+}/O^{6+}$ compatible with \textit{in situ} data are reproduced,
there are no solutions
than can simultaneously match the measured  $C^{6+}/C^{5+}$  and $O^{7+}/O^{6+}$. 
Panel (c) shows some superposition between 
measurements and calculations with the highest values of $z_+^\odot$. However, as it appears from
 Fig.~2 of \citet{2014ApJ...784..120L},  these large $z_+^\odot$ yields
plasma parameters at  1 A.U.\ that are not generally observed in the solar wind. On the contrary, the thin area in
gold in Panel (c), which corresponds to solutions with  $630 \lesssim U \lesssim 820 \ \mathrm{km/s}$ and
$1.5 \lesssim n_e \lesssim 3 \ \mathrm{cm^{-3}}$, does not intersect the bulk of Ulysses measurements.
\subsection{Latitudinal Profiles of Charge-States in the Solar Wind}  \label{ss:bana}
 \begin{figure} 
 \centerline{\includegraphics[width=\textwidth,clip]{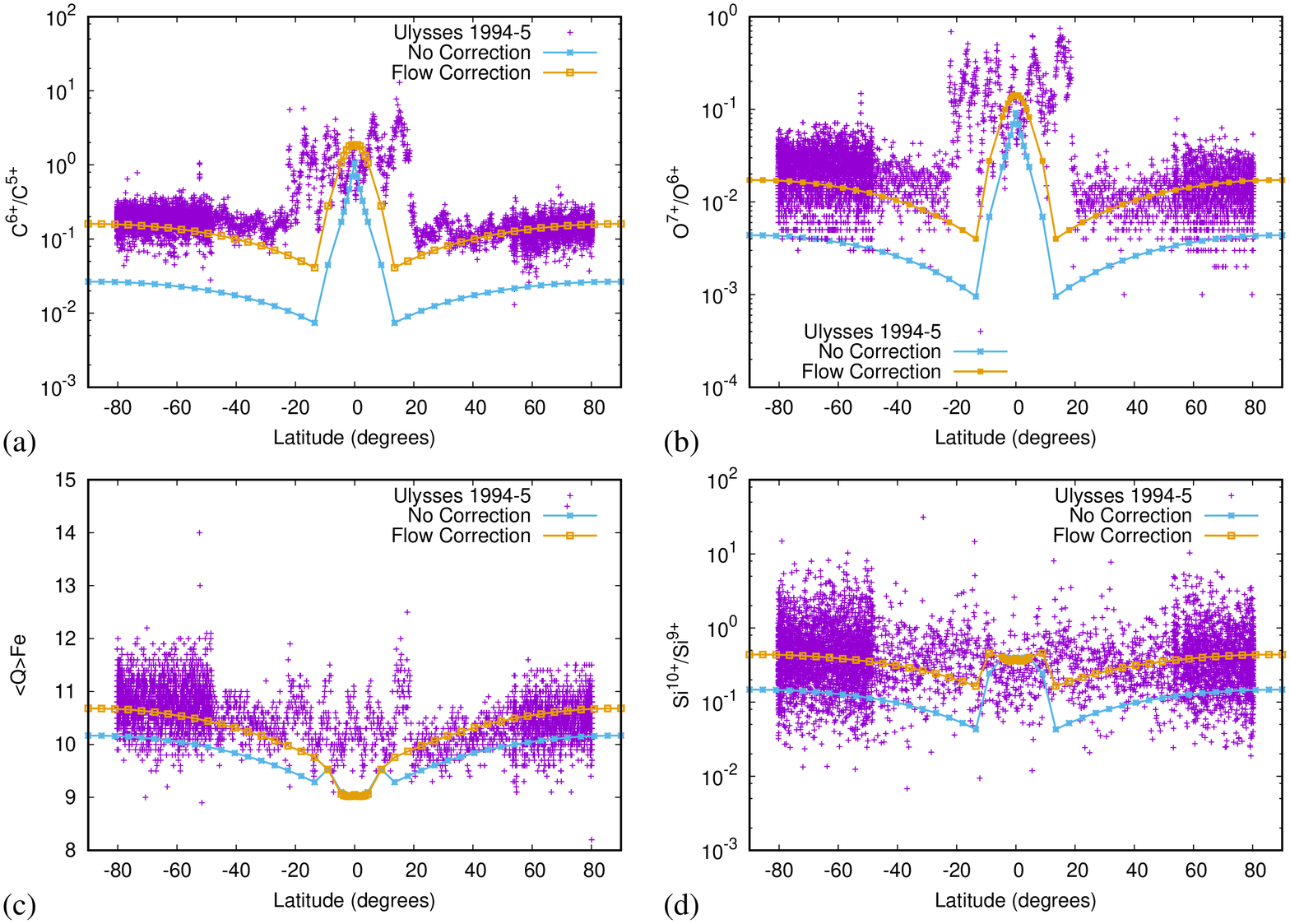}}
 \caption{Latitudinal dependence of ion charge states of the solar wind at 1 A.U.\ in WTD model of 
\citet{2014ApJ...796..111L} compared with the measurements of Ulysses  during 1994-5. {25 solar wind solutions (indicated with symbols along the curves) were calculated at latitudes 
between $0^\circ$ and $90^\circ$
 along field lines of the symmetric model of
 \citet{1998A&A...337..940B}.}
The cyan curves are for the unmodified charge states evolution model, the orange curves show the results when
a correction to the flow is applied.
(a)  $C^{6+}/C^{5+}$.
(b)  $O^{7+}/O^{6+}$.  
(c) $<Q>Fe$.
(d) $Si^{10+}/Si^{9+}$.
}\label{fig:bana}
 \end{figure}
{
\citet{2014ApJ...796..111L} used the WTD model of Sec.~\ref{s:model} to 
calculate solar wind solutions  along 25  magnetic  field lines 
 extracted at different latitudes between $0^\circ$ and $90^\circ$
 from the 2D, axisymmetric, 
analytic model of \citet{1998A&A...337..940B}. 
For each flux-tube, the same
combination of turbulence parameters $z_+^\odot = 54~\mathrm{km/s}$ and  $\lambda_\odot = 0.02~R_\odot$ was
employed.} The computed latitudinal dependence at 1 A.U.\ of plasma wind speed, number density, temperature, and pressure
 \citep[Fig.~2 of][]{2014ApJ...796..111L} was found to be in qualitative agreement with  the more advanced model of 
\citet{2007ApJS..171..520C} and within the range of \textit{in situ} data.

 We have also repeated 
the simulations of \citet{2014ApJ...796..111L} to calculate the charge states   for carbon, oxygen,  and iron.
In Fig.~\ref{fig:bana}a we compare the latitudinal dependence of the computed $C^{6+}/C^{5+}$ ratio (in cyan{, each symbol representing a solution}) with that 
measured by Ulysses during 1994-5. Although we cannot expect agreement
at low latitudes, where the charge states 
are affected  by  the properties of
the equatorial streamer and possible encounters with CMEs,  the results of the simulations are about one order of magnitude too low even at the poles.
The  calculated  $O^{7+}/O^{6+}$ ratios (in cyan) in Fig.~\ref{fig:bana}b are also too low. The curve of simulated 
average iron charge states,  $<Q>Fe$, 
which is   depicted in cyan in Fig.~\ref{fig:bana}c, 
is at the lower limit of the measurements.
\subsection{Correcting the Ion Outflow Speed} 
 \begin{figure} 
 \centerline{\includegraphics[width=\textwidth,clip]{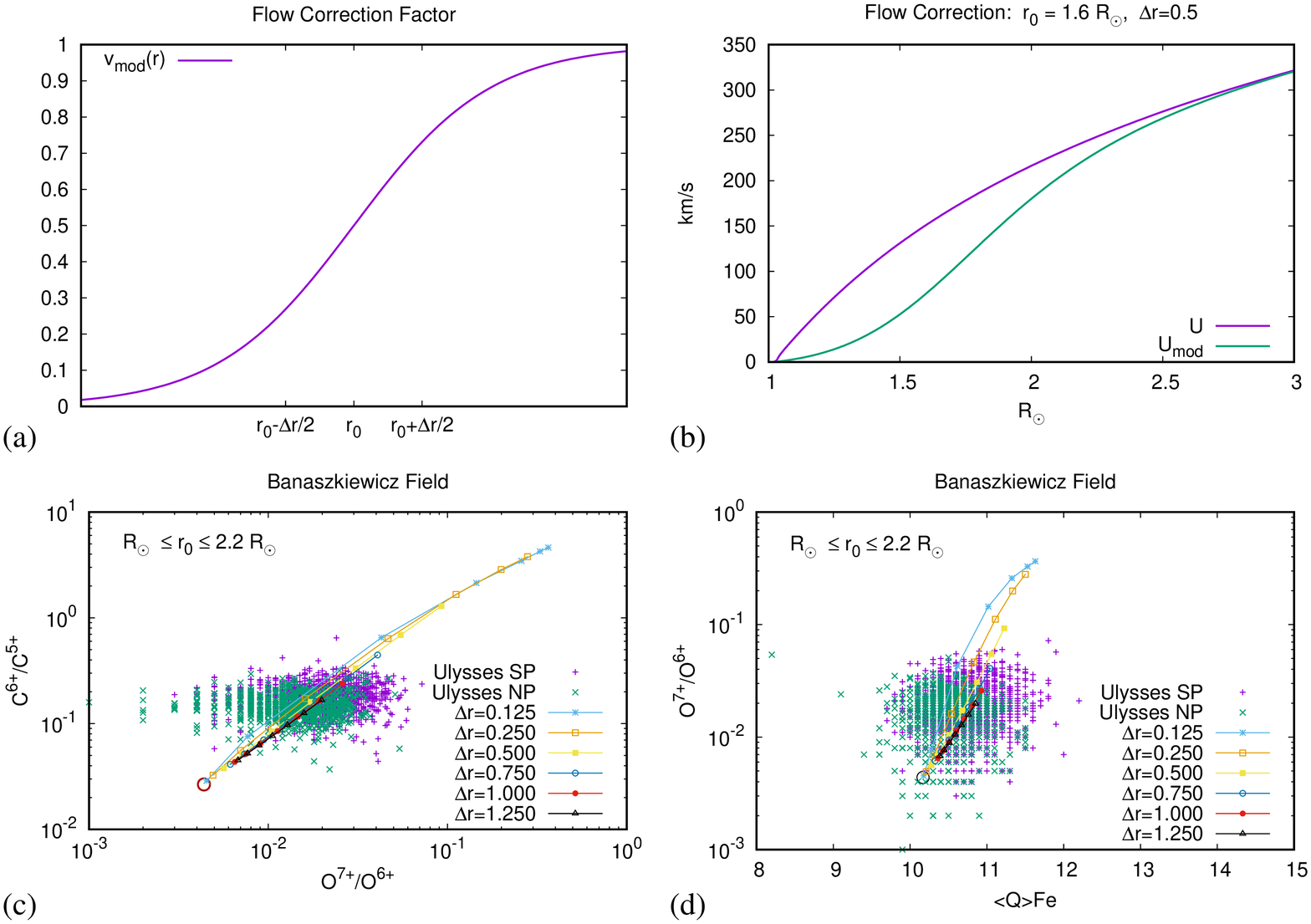}}
 \caption{
(a) A plot of  the two-parameter (i.e.,  $\Delta r$ and $r_0$) function in Eq.~(\ref{e:vmod}).
(b) In purple, the solar wind speed  along  the polar magnetic field line of the model
of \citet{1998A&A...337..940B}  calculated with the WTD algorithm.
In green, the speed used to advance the charge states in Eq.~(\ref{e:cs}) when $v_\mathrm{mod}(r)$  
with  $\Delta r=0.5 R_\odot$ and $r_0= 1.6 R_\odot$ is applied to the flow as in Eq.~(\ref{e:umod}).
(c)  $O^{7+}/O^{6+}$ vs.\ $C^{6+}/C^{5+}$  at 1 A.U.\ for the same field line. The circle in the lower left corner
 shows the values if no correction is applied to the flow in  Eq.~(\ref{e:cs}). 
Values along each curves, from bottom-left to top-right,
 represents solutions with the same 
 $\Delta r$ in $v_\mathrm{mod}(r)$ and increasing $r_0$, from 
1 to $2.2~R_\odot$ at intervals of $0.2~R_\odot$.
The curves are superimposed to the Ulysses measurements at latitudes 
$|\phi| \geq  70^\circ$ during 1994-5.
(d)  The same as (c) but for $<Q>Fe$  vs.\ $O^{7+}/O^{6+}$.
}\label{fig:varying}
 \end{figure}
Since the results in Subsecs.~\ref{ss:fastwind} and \ref{ss:bana} show that the WTD model described in 
Sec.~\ref{s:model} cannot reproduce the charge states of ions in the solar wind, we have looked for possible improvements
that may also be implemented in the 3D model of \citet{2018NatAs.tmp..120M}. One possible reason why the charge states
in our model are too low is that we do not include the effect of a suprathermal electron tail in the corona that would
increase the ionization coefficients
  \citep{1996GeoRL..23.2785K,1998ApJ...498..448E,2014ApJ...791L..31C}. However, no conclusive evidence of such non-Maxwellian distribution has yet
emerged \citep{2009LRSP....6....3C}.  Another possibility is that a simple, one fluid model does not account for the possibility 
that ions traveling  at lower speeds than electrons would 
spend more time in the lower corona,
where they would likely reach higher charge states \citep{1998JGR...10314539K}. 
\citet{2014ApJ...790..111L}
 proposed a correction to the flow in the model of \citet{2007ApJS..171..520C} 
{to have the
source region located in the corona rather than in the lower atmosphere. 
The charge states of the solar wind were already
 closer to the measured ones,
but still a better agreement was reached mostly due to this fact.  }
 Inspired by their work, 
we intend to determine a modifying
 factor $v_\mathrm{mod}(r)$ such that when applied to $U(s)$,
\begin{equation}
U_\mathrm{mod}(s)=v_\mathrm{mod}(r)U(s), \label{e:umod}
\end{equation}
may give a smaller ion outflow speed in the lower corona,
and thus make  the
 charge states at 1 A.U.\ as calculated in Eq.~(\ref{e:cs}) 
 compatible with the Ulysses measurements.
We choose for $v_\mathrm{mod}(r)$ the following formulation:
\begin{equation}
v_\mathrm{mod}(r)=\frac{1}{2} \left ( 1+ \tanh \frac{r-r_0}{\Delta r} \right ), \label{e:vmod}
\end{equation}
which is based on two parameters, $r_0$ and $\Delta r$.  As Fig.~\ref{fig:varying}a shows, $r_0$ controls where the flow is switched on 
 and  $\Delta r$ is the interval over which this transition occurs.
To determine heuristically the optimal values of these parameters,
we repeat the charge-states calculation
for the polar field line in Subsec.~\ref{ss:bana} with  $v_\mathrm{mod}$
having $r_0 = 1$, 1.2, 1.4, 1.6, 1.8, 2., $ 2.2~R_\odot$ 
and $\Delta r =0.125$, 0.25, 0.75, 1, $1.25~R_\odot$. Then we evaluate
for each solutions the values of the ratios $O^{7+}/O^{6+}$ 
and $C^{6+}/C^{5+}$ as well as $<Q>Fe$. Finally 
we select the couple $(r_0,\Delta r)$ that yields the results closest
to  the Ulysses measurements. Figures~\ref{fig:varying}c and \ref{fig:varying}d
show the calculated charge states respectively in the 
 $C^{6+}/C^{5+}$ vs. $O^{7+}/O^{6+}$ and in the  $<Q>Fe$ vs. 
$O^{7+}/O^{6+}$ planes.
Each curve corresponds to a given  $\Delta r$. The symbols along each
curve represent values of $r_0$ increasing from the bottom left (where 
the charge states
for the solution with no  $v_\mathrm{mod}$ are indicated with
circles) to top right. The values corresponding to the couple
$(r_0=1.6~R_\odot ,\Delta r=0.5 R_\odot)$ fall close to the centers of the
Ulysses measurements. With this choice, the ion outflow speed is modified
as depicted in Fig.~\ref{fig:varying}b.
Since the ions are traveling for a longer time in
the lower corona, they can reach
higher charge states  before  being ``frozen-in."
\subsection{Charge States Calculations with Modified Ion Outflow Speed}
We have repeated the calculations of Subsec.~\ref{ss:fastwind} with the
ion outflow speed
 modified with  $v_\mathrm{mod}$ according to the optimal choice
of parameters ($\Delta r=0.5 R_\odot$ and $r_0= 1.6 R_\odot$) as
described in the previous subsection. The effects of the modification
can be seen in Figures~\ref{fig:fastwind}b and \ref{fig:fastwind}d, which
are the respective counterparts of the unmodified calculations in
Figs.~\ref{fig:fastwind}a and \ref{fig:fastwind}c.  Higher charge states
are reached so that 
now the thin areas in
gold, which correspond to solar wind
 solutions with  $630 \lesssim U \lesssim
 820 \ \mathrm{km/s}$ and
$1.5 \lesssim n_e \lesssim 3 \ \mathrm{cm^{-3}}$, intersect (or at least
touch) the bulk of
 Ulysses measurements.
Hence, solutions in these subregions have not only plasma velocity
and density, but also charge states values
 compatible with \textit{in situ} measurements.

Analogously, we  have recalculated the latitudinal profiles of
Subsec.~\ref{ss:bana} applying $v_\mathrm{mod}$ to slow down the 
flow of ions. The orange curves in panels a, b, and c of Fig.~\ref{fig:bana},
which 
correspond to simulations with  the corrected ion outflow speed,
show  higher charge states being formed in comparison with the
cyan curves, for which no such modification is applied. Thus, at least
for the higher latitudes, the calculated charge states
 lie now close to the middle of the bulk of the
Ulysses measurements. 

To verify our approach, we  also calculate the  $Si^{10+}/Si^{9+}$
ratio, which has
  not been used to optimize the parameters of the $v_\mathrm{mod}$ 
function {and for which there exists data in the Ulysses/SWICS archive}. 
The resulting 
latitudinal profiles, with and without the ion outflow speed
correction, are  shown in Fig.~\ref{fig:bana}d {superimposed 
to the  measurements. These, due to uncertainties,
span about two orders of magnitude. Although
both curves fall within the bulk of the data,  
the profile with 
flow correction lies  closer to the average value. This confirms that
 our approach is not, al least, inferior to that using the unmodified flow.}

\section{Conclusions}\label{s:conclusion} 
We have  incorporated time-dependent fractional charge states evolution
into our 1D WTD model of the solar wind. We have implemented this
capability
with the aim of introducing  it
also into our 3D MHD model of the solar corona
and inner heliosphere. In fact,  charge states calculations, especially when
combined with other EUV, X-ray, and white light emission diagnostics, 
represent a powerful constraint  on the underlying WTD MHD model.
They can provide additional constraints on the
correlation scale of the turbulence 
and the amplitude of the outwardly propagating Alfv\'en perturbation 
at the solar surface.
However, the charge states percentages as calculated from the WTD model
do not match  the heliospheric measurements 
taken by Ulysses in 1994-5. We have
heuristically determined a correction to the ion outflow speed
 to be used to evolve the  charge states.
This yields, particularly for the polar regions, a 
better agreement between the calculated values and
the   \textit{in situ} measurements 
of Ulysses during 1994-5. At lower latitudes, where there are  uncertainties
due to possible encounters with  CMEs and the  configuration of the 
equatorial streamer, the discrepancy is larger.
{Comparing our work with that of \citet{2015ApJ...806...55O},
we notice first the differences between
 their models and ours: \citeauthor{2015ApJ...806...55O} employed a 
global MHD algorithm
driven by a sophisticated Alfv\'en WTD formulation, selected field lines
at different latitudes,  used an external code to evaluate the charge states
along the same, and compared the results with the measurements of Ulysses
during its third polar scan of 2007. Yet, despite all these differences,
their models  disagreed with measurements in the same sense as ours, namely
 ionization rates were underpredicted.  
 \citet{2015ApJ...806...55O} considered the same explanations we discuss
in the text, but finally
invoked suprathermal electrons as a possible, unaccounted
 mechanism to close the gap with observations. We have
postulated a slower propagation speed for the ions.
}
Although the ion outflow speed modification, which
was inspired by that of \citet{2014ApJ...790..111L}, 
may capture some of the physics of the ions, 
there are
several other possible explanations for the mismatch between the 
calculated charge states
and  the observations. 
 On the other hand,
our modified ion outflow speed 
(Fig.~\ref{fig:varying}b) lies within the range of recent empirical results
\citep[Fig.~4 of][]{2016SSRv..201...55A},
since it is already more than 300 km/s at $3~R_\odot$.  It is also
possible that photoionization may yield the
 higher charge states  measured in the solar wind{. Even if 
 \citet{2015ApJ...812L..28L} found that it could be
a significant factor, yet  it was not sufficient 
 to explain the discrepancy between predictions and measurements.} 
We plan to explore this effect in future work.
{Moreover, the plasma density and temperature in the lower
corona could also be  factors of critical importance
in setting the charge state distribution of the solar wind.
Although our model was shown to provide results compatible with
observations \citep{2014ApJ...796..111L}, we cannot categorically
exclude 
that a different heating model could yield not only the same
plasma parameters at 1 AU, but also conditions in the lower
corona causing higher ionization.
}
Needless to say,
a more accurate calculation of charge states would also require 
multi-fluid simulations \citep[e.g.,][]{2013ApJ...762...18O} or even
  multi-ions simulations  \citep[e.g.,][]{2011ApJ...732..119B}. 
In particular, as Fig.~3b of \citet{2011ApJ...732..119B} shows, 
a single outflow speed for all ions is only a
crude approximation. Introducing a more realistic evolution of the plasma,
starting from evolving the temperature of electrons and protons
separately, is a first step into this direction that will be implemented
next. However, considering the end goal of our investigation is
to provide accurate 3D modeling of the corona and heliosphere capable of
predicting tomorrow's conditions from today's empirical
 data, compromises on which physical mechanisms to include next will
be inevitable.
 They will also be acceptable only if the
results can be quantitatively matched with observations.

%

%

%

%
 \begin{acks}
RL is grateful to Drs.\ Susanna Parenti	 and
Alessandro Bemporad for providing helpful advice. RL was funded through
NASA Grant 
NNH14CK98C.
 \end{acks}

%
%
\bibliographystyle{spr-mp-sola}
\bibliography{mybib}  
%
%
%
%

\end{article} 
\end{document}